\documentclass[12pt,epsf]{article}
\usepackage{amssymb,amsmath,amsbsy}
\usepackage{graphicx}

\newcommand{\be}{\begin{equation}}
\newcommand{\ee}{\end{equation}}
\newcommand{\bea}{\begin{eqnarray}}
\newcommand{\eea}{\end{eqnarray}}
\newcommand{\bear}{\begin{eqnarray}}
\newcommand{\eear}{\end{eqnarray}}
\newcommand{\beas}{\begin{eqnarray*}}
\newcommand{\p}{\partial}
\newcommand{\eeas}{\end{eqnarray*}}
\newcommand{\ba}{\begin{array}}
\newcommand{\ea}{\end{array}}

\renewcommand{\l}{\lambda}

\newcommand{\tr}{\operatorname{tr}}
\newcommand{\pd}[2][1]{\ifnum#1=1 \frac{\partial}{\partial {#2}} \else
  \frac{\partial^#1}{\partial {#2}^{#1}}\fi}
\newcommand{\dpd}[2][1]{\ifnum#1=1 \dfrac{\partial}{\partial {#2}} \else
  \frac{\partial^#1}{\partial {#2}^{#1}}\fi}
\newcommand{\td}[2][1]{\ifnum#1=1 \frac{d}{d{#2}} \else
  \frac{d^#1}{d{#2}^{#1}}\fi}

\newcommand{\nbox}{{\,\lower0.9pt\vbox{\hrule \hbox{\vrule height 0.2 cm \hskip 0.19 cm \vrule height 0.2 cm}\hrule}\,}}

\def\href#1#2{#2}

\begin{document}
\begin{titlepage}
\hfill
\vbox{
    \halign{#\hfil         \cr
           } 
      }  
\vspace*{20mm}
\begin{center}
{\Large \bf Canonical Energy is Quantum Fisher Information}

\vspace*{15mm}
\vspace*{1mm}
Nima Lashkari$^{a,b}$ and Mark Van Raamsdonk$^b$
\vspace*{1cm}
\let\thefootnote\relax\footnote{lashkari@mit.edu, mav@phas.ubc.ca}

{${}^a$ Center for Theoretical Physics, Massachusetts Institute of Technology\\
77 Massachusetts Avenue, Cambridge, MA, 02139, USA\\
\vspace*{0.2cm}
${}^{b}$ Department of Physics and Astronomy,
University of British Columbia\\
6224 Agricultural Road,
Vancouver, B.C., V6T 1W9, Canada}

\vspace*{1cm}
\end{center}
\begin{abstract}

In quantum information theory, Fisher Information is a natural metric on the space of perturbations to a density matrix, defined by calculating the relative entropy with the unperturbed state at quadratic order in perturbations. In gravitational physics, Canonical Energy defines a natural metric on the space of perturbations to spacetimes with a Killing horizon. In this paper, we show that the Fisher information metric for perturbations to the vacuum density matrix of a ball-shaped region $B$ in a holographic CFT is dual to the canonical energy metric for perturbations to a corresponding Rindler wedge $R_B$ of Anti-de-Sitter space. Positivity of relative entropy at second order implies that the Fisher information metric is positive definite. Thus, for physical perturbations to anti-de-Sitter spacetime, the canonical energy associated to any Rindler wedge must be positive. This second-order constraint on the metric extends the first order result from relative entropy positivity that physical perturbations must satisfy the linearized Einstein's equations.

\end{abstract}

\end{titlepage}

\vskip 1cm

\section{Introduction}

In the AdS/CFT correspondence \cite{maldacena1997large}, the holographic entanglement entropy formula \cite{ryu2006holographic,hubeny2007covariant} relates the entanglement structure of the CFT with the geometrical structure of the dual spacetime. On the CFT side, the entanglement structure obeys fundamental consistency constraints such as the strong subadditivity of entanglement entropy and the positivity and monotonicity of relative entropy.\footnote{For a review, see for example \cite{nielsen2010quantum}.} These translate to geometrical constraints that must be satisfied for geometries dual to consistent CFT states \cite{Callan:2012ip,blanco2013relative,Lin:2014hva,Lashkari:2014kda,Bhattacharya:2014vja}. To leading order in perturbations away from the vacuum state, these constraints (specifically the positivity of relative entropy) translate to the statement that the dual geometry must satisfy Einstein's equations to linear order in perturbations around AdS \cite{lashkari2014gravitational,faulkner2014gravitation,swingle2014universality} (see also \cite{Faulkner:2014jva}). In this paper, we extend this work to give a complete characterization of the positivity of relative entropy constraints to second order in perturbations to the vacuum. We have a constraint for each ball-shaped region $B$ in the CFT; these constraints imply the positivity ``canonical energy,'' a quantity quadratic in the metric perturbations to a Rindler wedge region $R_B$ associated with $B$. The results in this paper make use of an important identity in classical theories of gravity relating the gravity dual of relative entropy to the natural symplectic form on the space of perturbations to a metric \cite{hollands2013stability}.

We now present a concise summary of the background and results before giving an outline of the remainder of the paper.

\subsubsection*{Fisher Information in conformal field theory}

Consider a one-parameter family of states $|\Psi(\lambda) \rangle$ of a CFT on $R^{d-1,1}$ with $| \Psi(0) \rangle$ the vacuum state. For any ball-shaped region $B$, define $\rho_B (\lambda)$ as the reduced density matrix for this region. We have that
\[
\rho_B(0) = {1 \over Z} e^{- H_B}
\]
where $H_B$ (the modular Hamiltonian for the subsystem $B$ in the vacuum state) is the generator of a conformal Killing vector $\zeta_B$ acting in the causal diamond region $D_B$ associated with $B$, as shown in figure 1.\footnote{Explicitly, we have
\be
\label{defzeta}
\zeta_B = \frac{\pi}{R}( \left[R^2  - (t-t_0)^2 + |\vec{x}-\vec{x}_0|^2\right] \partial_t - \left[2 (t-t_0)(x^i -
x_0^i)\right]\partial_i)
\ee for the ball of radius $R$ centered at $(t_0,x_0^i)$.}
For a ball of radius $R$, we have \cite{casini2011towards}
\[
H_B = 2 \pi \int d^{d-1} x {R^2 - r^2 \over 2 R} T_{00} \; .
\]
where $r$ is the distance to the center of the ball. For any state $|\Psi(\lambda) \rangle$ we define
\[
\Delta S_B = S(\rho_B(\lambda)) - S(\rho_B(0))
\]
as the difference in entanglement entropy compared with the vacuum state. We also define
\[
\Delta E_B = \tr(H_B \rho_B(\lambda)) - \tr(H_B \rho_B(0))
\]
as the difference in the expectation value of the modular Hamiltonian. Both $\Delta S_B$ and $\Delta E_B$ are finite for well-behaved states. Positivity of relative entropy (reviewed below) gives the fundamental constraint that \cite{blanco2013relative}
\be
\label{RE1}
\Delta E_B(\lambda) - \Delta S_B(\lambda) \ge 0 \; .
\ee
Since $\lambda=0$ represents a minimum for any family of perturbations, we must have
\be
\label{firstlaw}
{d \over d \lambda}( \Delta E_B(\lambda) - \Delta S_B(\lambda))|_{\lambda = 0} = 0 \; ,
\ee
known as the first law of entanglement \cite{blanco2013relative}. At second order in $\lambda$, the constraint becomes
\be
\label{RE2}
{d^2 \over d \lambda^2}( \Delta E_B(\lambda) - \Delta S_B(\lambda))|_{\lambda = 0} \ge 0 \; .
\ee
The quantity on the left here defines Fisher Information. It is a quadratic form $\langle \delta \rho_B, \delta \rho_B \rangle_{\rho_B(0)}$ in the first order perturbation $\delta \rho_B = \partial_\lambda \rho_B|_{\lambda=0}$ to the unperturbed state. This can be promoted to a metric on perturbations
\be
\label{Fishermetric}
\langle \delta \rho, \delta \sigma \rangle_{\rho(0)} \equiv {1 \over 2} (\langle \delta \rho + \delta \sigma, \delta \rho + \delta \sigma \rangle - \langle \delta \rho , \delta \rho \rangle - \langle \delta \sigma, \delta \sigma \rangle) \; .
\ee
The second order statement of positivity of relative entropy is thus that Fisher Information metric is positive definite. The Fisher information of perturbations near vacuum in conformal field theory for ball-shaped regions is known to be related to 2-point function of the theory and universal \cite{Lashkari:2015dia}.

\begin{figure}
\centering
\includegraphics[width=0.2\textwidth]{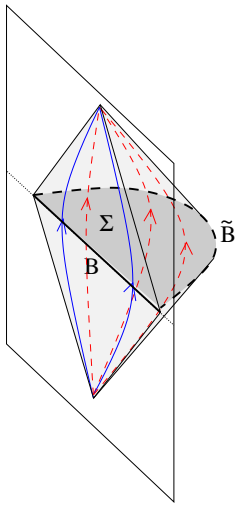}\\
\caption{AdS-Rindler wedge $R_B$ associated with a ball $B$ on a spatial slice of the boundary. $R_B$ is the intersection of the causal past and the causal future of the domain of dependence $D_B$ (boundary diamond). Solid blue paths indicate the boundary flow associated with $H_B$ and the conformal Killing vector $\zeta$. Dashed red paths indicate the action of the Killing vector $\xi$.}
\label{hyperbolic2}
\end{figure}

\subsubsection*{Gravity interpretation}

Now suppose that the CFT is holographic and that the one-parameter family of states $|\Psi(\lambda)\rangle$ have gravity dual geometries $M(\lambda)$ with $M(0)$ equal to pure $AdS$.
In this unperturbed geometry, the ball shaped-region $B$ can be associated \cite{Czech:2012be} with a Rindler wedge $R_B$ defined as the intersection of the causal past and the causal future of $D_B$, the boundary domain of dependence of $B$ (see figure \ref{hyperbolic2}) (see also \cite{Bousso:2012sj,Czech:2012bh,Hubeny:2012wa}). The boundary of this Rindler wedge is the extremal area surface $\tilde{B}$ in the bulk with boundary $\partial \tilde{B} = \partial B$. By a change of coordinates, the wedge $R_B$ is seen to be diffeomorphic to the exterior of a hyperbolic Schwarzchild-AdS black hole for which $\tilde{B}$ is the horizon.\footnote{This is related to the field theory statement that a conformal transformation maps the region $D_B$ to hyperbolic space times time, mapping the vacuum density matrix on $D_B$ to the $T = 1/(2 \pi R_H)$ thermal state on hyperbolic space with curvature radius $R_H$ \cite{casini2011towards}.} The wedge $R_B$ has a timelike Killing vector $\xi_B$ vanishing on $\tilde{B}$ that extends $\zeta_B$ into the bulk and defines a ``Rinder time'' for the wedge.

For the perturbed asymptotically AdS dual geometry $M(\lambda)$ we can define $\tilde{B}(\lambda)$ to be the extremal surface in $M(\lambda)$ with $\partial \tilde{B}(\lambda) = \partial B$. We can define $R_B(\lambda)$ to be the set of points in $\tilde{M}$ that are spacelike separated from $\tilde{B}$ towards the boundary \cite{Czech:2012bh,Wall:2012uf}. Thus, as we deform the CFT state, each wedge $R_B$ is deformed to $R_B(\lambda)$ that can be viewed as a perturbed hyperbolic black hole. Using the holographic entanglement entropy formula, the CFT quantity $\Delta S_B$ corresponds to the change in area of $\tilde{B}$ as the geometry is varied from $M(0)$ to $M(\lambda)$. As we review below, there is a natural gravitational energy $E^{grav}_B$, calculated from the asymptotic metric near $B$, that can be associated with any $R_B(\lambda)$ \cite{Iyer:1994ys}. The field theory quantity $\Delta E_B$ is related to the change in gravitational energy for $R_B$ as the geometry is varied from $M(0)$ to $M(\lambda)$.

We can now translate the relative entropy constraint (\ref{RE1}) to a gravitational statement. For any geometry $M(\lambda)$ dual to a physical CFT state, we must have \cite{blanco2013relative}
\be
\label{firstdiff}
\Delta E^{grav}_B - \Delta S_B^{grav} \ge 0 \; .
\ee
Thus, for every ball $B$, the change in area of the extremal surface $\tilde{B}$ is bounded by the change in gravitational energy for the region $R_B$. At first order, according to (\ref{firstlaw}), these changes must be equal, so we have a gravitational first law
\[
\delta E_B^{grav} = \delta S_B^{grav} \;
\]
governing perturbations of hyperbolic black holes. The recent work of \cite{lashkari2014gravitational} shows that the collection of these first law statements for all $B$ is equivalent to a single local bulk constraint, that the first order perturbation satisfies the linearized Einstein equation.

A powerful method \cite{faulkner2014gravitation} to prove this first order result makes use of a gravitational identity of Wald and Iyer \cite{Iyer:1994ys} relating the difference (\ref{firstdiff}) to the integral of a bulk quantity over a surface $\Sigma_B$ bounded by $B$ and $\tilde{B}$:
\be
\label{WI}
{d \over d \lambda} (\Delta E^{grav}_B - \Delta S_B^{grav})|_{\lambda=0} = \int_{\Sigma_B} \hat{E}_B(\delta g) \; .
\ee
Here, $\hat{E}_B$ is a form that vanishes when the metric perturbation $\delta g$ satisfies the linearized Einstein equations. Since the field theory result (\ref{firstlaw}) implies the vanishing of the left side here, we immediately have that all $\hat{E}_B$ integrals vanish. It is straightforward to show that this is impossible unless the metric perturbation satisfies the linearized Einstein equations.

The key technical tool in this paper is a result by Hollands and Wald \cite{hollands2013stability} generalizing the gravitational identity (\ref{WI}) away from $\lambda=0$. The full result takes the form
\be
\label{WH}
{d \over d \lambda} (\Delta E^{grav}_B - \Delta S_B^{grav}) = W_{\Sigma_B}(g(\lambda), \partial_\lambda g(\lambda)) +  \int_{\Sigma_B} \hat{E}_B(g(\lambda),\partial_\lambda g(\lambda))
\ee
where again $\hat{E}_B$ is a quantity that vanishes when the metrics $g(\lambda)$ are on shell (i.e. satisfy the nonlinear gravitational equations), and $W_{\Sigma_B}$ is another integral over $\Sigma_B$ defined in terms of a natural symplectic form defined on the space of perturbations to the metric $g(\lambda)$. The identity (\ref{WH}) allows us to rewrite the difference of boundary integrals defining the gravity dual of relative entropy (left side of (\ref{WH})) as the integral over a bulk quantity. Specializing to the terms in (\ref{WH}) at order $\lambda$ (i.e. the $\lambda$ derivative of (\ref{WH}) at $\lambda=0$), the result reduces to
\be
\label{Fisher2}
{d^2 \over d \lambda^2} (\Delta E^{grav}_B - \Delta S_B^{grav})_{\lambda=0} = {\cal E}_B(\delta g, \delta g)
\ee
where ${\cal E}_B(\delta g, \delta g)$ is a quadratic form on the metric perturbations known as ``canonical energy.'' Essentially, it is the Rindler energy (associated with the Killing vector $\xi_B$) for the Rindler wedge $R_B$, including a gravitational piece (quadratic in the metric perturbation) and a matter contribution:\footnote{Here, the gravitational contribution implicitly includes a term involving an integral over $\tilde{B}$, as described in section 3.}
\be
\label{canonical}
{\cal E}(\gamma, \gamma) = \int \xi^a (T^{grav}_{ab} + T^{matter}_{ab}) d \Sigma^b \; .
\ee
The left side of (\ref{Fisher2}) is exactly the gravity dual of Fisher information (\ref{RE2}). So we have that Fisher Information is dual to canonical energy. Consequently, the positivity of Fisher information translates to the positivity of the canonical energy ${\cal E}_B$ for each Rindler wedge $B$. Here, the ``matter'' contribution is actually $T^{matter}_{ab} = {1 \over 8 \pi} (G_{ab} - \Lambda g_{ab})$, so this is a purely geometrical constraint, but we can rewrite this as the matter stress tensor assuming that Einstein's equations are satisfied. In this case the positivity of (\ref{canonical}) can be interpreted as an energy condition restricting the behavior of the matter stress tensor in a consistent theory.

It is quite natural that Fisher information and canonical energy are related to one another, since each defines a natural metric on a space of perturbations, in one case to a density matrix, and in the other case to a metric satisfying the gravitational equations. This identification provides is further evidence that the geometry of spacetime in quantum gravity is fundamentally related to the entanglement structure of the fundamental degrees of freedom.

\subsubsection*{Outline}

The remainder of this paper is organized as follows. In section 2, we provide in more detail the background material on relative entropy, quantum Fisher information, and the tools to translate these to dual geometrical quantities in holographic theories. In section 3, we review the fundamental gravitational identity of Hollands and Wald that allows us to translate the gravitational expression dual to relative entropy (which can be expressed as a boundary integral over the surface $B - \tilde{B}$) to a bulk quantity. We review the definition of canonical energy and show that this provides the gravity dual of quantum Fisher information. Finally, we express the positivity of Fisher information as an explicit constraint on the dual geometry, showing that it may be written in the form of an energy condition that must be obeyed by the matter stress tensor. In section 4 we provide some example calculations, discussing in general how to calculate canonical energy for an on-shell metric perturbation given in a general gauge, and providing some explicit example calculations in $AdS_3$. These calculation give explicit constraints on the second order metric for physical asymptotically $AdS_3$ geometries. We check in particular that the constraints on the asymptotic metric exactly reproduce those calculated previously in \cite{Lashkari:2014kda}. We conclude in section 5 with a discussion.

Note added: While this manuscript was in preparation, the paper \cite{MIyaji:2015mia} appeared, which discusses the gravitational interpretation of a different type of quantum information metric. The metric discussed there is defined in terms of the inner product between states rather than the relative entropy between states, and the proposed gravity dual in \cite{MIyaji:2015mia} involves the volume of a spatial slice rather than the canonical energy. Thus, the two papers represent two independent elements in the quantum information / quantum gravity dictionary.

\section{Background}

In this section, we review in more detail the definition of relative entropy and its positivity and monotonicity properties, starting from general quantum systems, and then specializing to the case of conformal field theories. We then recall how the quantities entering into the formula for relative entropy are related to gravitational quantities in the case of holographic CFTs.

\subsection{Relative Entropy and Fisher Information}

Relative entropy measures the distinguishability of a density matrix $\rho$ from some reference density matrix $\sigma$. It is defined as
\[
S(\rho || \sigma) = \tr(\rho \log \rho) - \tr(\rho \log \sigma) \; .
\]
The relative entropy is always nonnegative, equal to zero for identical states and increasing to infinity if $\rho$ has nonzero probability for a state orthogonal to the subspace of states in the ensemble described by $\sigma$. Further, relative entropy is {\it monotonic}: if $A$ represents a subsystem of some quantum system $B$, and if $\rho_A$ and $\sigma_A$ are the reduced density matrices for the subsystem obtained from $\rho_B$ and $\sigma_B$, then
\[
S(\rho_A || \sigma_A) \le S(\rho_B || \sigma_B) \; .
\]
Detailed proofs of these results may be found in \cite{nielsen2010quantum}.

These results are particularly useful when the density matrix for the reference state is known explicitly. In this case, defining the modular Hamiltonian
\[
H_\sigma = - \log(\sigma) \; ,
\]
we have
\bea\label{deltaF}
S(\rho || \sigma) &=& \tr(\rho \log \rho) - \tr(\sigma \log \sigma)  +  \tr(\sigma \log \sigma) - \tr(\rho \log \sigma) \cr
&=& \langle - \log \sigma \rangle_\rho - \langle - \log \sigma \rangle_\sigma  - S(\rho) + S(\sigma) \cr
&=& \Delta  \langle H_\sigma \rangle - \Delta S.
\label{reldef}
\eea
In this paper, we will mostly be interested in the relative entropy for nearby states, considering a one-parameter family
\[
\rho(\lambda) = \rho_0 + \lambda \rho_1 + \lambda^2  \rho_2  + {\cal O}(\lambda^3) \; ,
\]
with $\rho_0 = \sigma$. To first order in $\lambda$, it is straightforward to check that the relative entropy vanishes, a result known as the ``first law of entanglement,'' \cite{blanco2013relative}
\[
\delta S = \delta  \langle H_\sigma \rangle \; .
\]
At the second order in $\lambda$, relative entropy is given by\footnote{Note that the terms involving $\rho_2$ vanish by the entanglement first law applied to the perturbation $\lambda^2 \rho_2$.}
\bea\label{fisher}
S(\rho(\lambda)||\rho_0)_{\lambda^2} = \lambda^2 \langle \rho_1, \rho_1 \rangle_{\rho_0},
\eea
where
\bea\label{firstinner}
\langle \delta \rho, \delta \rho \rangle_\sigma \equiv  \tr \left(\delta \rho \frac{d}{d\lambda}\log(\sigma+\lambda \delta \rho)\Big |_{\lambda=0} \right) + \tr \left( \sigma {1 \over 2} \frac{d^2}{d\lambda^2}\log(\sigma+\lambda \delta \rho)\Big |_{\lambda=0} \right) \; .
\eea
Note that for all $\lambda$ we have $\tr(\rho(\lambda)\p_\lambda \log\rho(\lambda))=\tr(\delta\rho)=0$. Taking a $\lambda$ derivative of this expression gives
\bea\label{tracelesseq}
\tr \left(\delta \rho \frac{d}{d\lambda}\log(\sigma+\lambda \delta \rho)\Big |_{\lambda=0} \right) + \tr \left( \sigma \frac{d^2}{d\lambda^2}\log(\sigma+\lambda \delta \rho)\Big |_{\lambda=0} \right)=0.
\eea
Plugging this back in (\ref{firstinner}) gives
\bea
\langle \delta \rho, \delta \rho \rangle_\sigma=\frac{1}{2}\tr \left(\delta \rho \frac{d}{d\lambda}\log(\sigma+\lambda \delta \rho)\Big |_{\lambda=0} \right) \; .
\eea

This quantity, a quadratic function of the first order perturbations, is known as {\it quantum Fisher information}. It can be promoted to an inner product on the tangent space to the manifold of states at $\sigma$ via (\ref{Fishermetric}). By the positivity of relative entropy, the quantum Fisher information is non-degenerate, non-negative and can be thought of as defining a Riemannian metric on the space of states.{\footnote{Using (\ref{tracelesseq}) it is straightforward to see that quantum Fisher information is symmetric in its arguments: $S(\sigma+\lambda\delta\rho\|\sigma)-S(\sigma\|\sigma+\lambda\delta\rho)=O(\lambda^3)$.}

Quantum Fisher information plays a central role in quantum state estimation which studies how to determine the density operator $\rho(\lambda)$ from measurements performed on $n$ copies of the quantum system \cite{petz2011introduction}.

\subsubsection{Relative entropy in conformal field theories}

In the rest of this paper, we focus on the case where our quantum system is a conformal field theory on $R^{d-1,1}$, our reference state is the CFT vacuum, and our subsystems are the fields in ball-shaped regions. In this case, the modular Hamiltonian corresponding to the reduced density matrix for a ball is \cite{blanco2013relative}
\be
\label{HmodBall}
H_B = 2 \pi \int_{|x|<R} d^{d-1} x \frac{R^2 - |x|^2}{2R}  T^{\text{CFT}}_{00} \; .
\ee
This may be obtained most easily by noting that the domain of dependence region of the ball can be mapped by a conformal transformation to a Rindler wedge of Minkowski space. The modular Hamiltonian for this Rindler wedge in the CFT vacuum state is well-known to be the Rindler Hamiltonian (boost generator), and the modular Hamiltonian (\ref{HmodBall}) is just the inverse conformal transformation applied to the Rindler Hamiltonian.

For a ball $B$, the relative entropy between the reduced density matrix $\rho_B$ in a general state and the vacuum density matrix $\sigma_B$ is then
\be
\label{defRE}
S(\rho_B || \sigma_B) = 2 \pi \int_{|x|<R} d^{d-1} x \frac{R^2 - |x|^2}{2R} \Delta \langle T^{\text{CFT}}_{00} \rangle - \Delta S_B \; .
\ee
Note that while relative entropy is well-defined for more general regions, it is only for ball-shaped regions that we can give an explicit form of the modular Hamiltonian as the integral of a local operator, and thus only in this case that we will be able to translate relative entropy to a gravitational quantity.

\subsection{Relative entropy in holographic conformal field theories}

We now consider the case of holographic conformal field theories for which the Ryu-Takayanagi formula \cite{ryu2006holographic} and its covariant generalization by Hubeny, Rangamani, and Takayanagi (HRT) \cite{hubeny2007covariant} holds. That is, we assume that there is a family of states $|\Psi \rangle$ and a related family of asymptotically AdS spacetimes $M_\Psi$ with boundary $R^{d-1,1}$  for which the entanglement entropy $S_A$ for any region $A$ is proportional to the area of the minimal area extremal surface $\tilde{A}$ in $M_\Psi$ for which $\partial A = \partial {\tilde{A}}$, where $A$ is the region on the boundary of $M_\Psi$ equivalent to the field theory region $A$. The proportionality constant is related to (or can be used to define) the gravitational Newton constant $G_N$ as
\be
\label{holoS}
S(A) = {{\rm Area}(\tilde{A}) \over 4 G_N} \equiv S^{grav} \; .
\ee

A useful explicit description of the spacetimes $M_\Psi$ is the metric in Fefferman-Graham coordinates, which takes the form
\be
\label{FGmetric}
ds^2 = {\ell_{AdS}^2 \over z^2}\left(dz^2 + dx_\mu dx^\mu + z^{d-1} \Gamma_{\mu \nu}(z,x) \right)
\ee
where $\Gamma_{\mu \nu}(z,x)$ has a finite limit as $z \to 0$.

With this assumption, we can compute the relative entropy of a holographic state $|\Psi \rangle$ using the dual geometry $M_\Psi$. The term $\Delta S$ is exactly the difference in area of the extremal surface $\tilde{A}$ in the geometry $M_\Psi$ compared with the geometry $M_{|vac\rangle} = AdS_{d+1}$. To calculate the term $\Delta \langle H_B \rangle$, we can use the fact that the HRT formula implies \cite{faulkner2014gravitation} that the CFT stress tensor expectation value is related to the asymptotic behavior of the metric (\ref{FGmetric}) as
\be
\label{holoT}
\Delta \langle T_{\mu \nu} \rangle = \Delta T_{\mu\nu}^{grav} \equiv {d \ell^{d-3} \over 16 \pi G_N}
\, \Gamma_{\mu \nu}(x,z=0)\ \ .
\ee
Using this, we have
\be
\label{holoE}
\Delta \langle H_B \rangle = {d \ell^{d-3} \over 8 G_N} \int_{|x|<R} d^{d-1} x \frac{R^2 - |x|^2}{2R}
\, \Gamma_{0 0}(x,z=0) \equiv \Delta E^{grav}
\ee
Thus, for holographic states, we have
\be
\label{REdict}
S(\rho_B || \rho^{vac}_B) = \Delta E_B - \Delta S_B = \Delta E^{grav}_B - \Delta S^{grav}_B
\ee
where $\Delta E^{grav}_B$ is defined by the boundary integral (\ref{holoE}) and $\Delta S^{grav}_B$ is defined via (\ref{holoS}) as the area difference for the extremal surface with boundary $\partial B$ between the geometries $M_\Psi$ and $M_{vac}$.

\section{Constraints on spacetime geometry from relative entropy inequalities}

For a holographic CFT state $|\Psi \rangle$ with a gravity dual geometry $M_\Psi$, equation (\ref{REdict}) provides a geometrical interpretation for the relative entropy with the vacuum state for a ball-shaped region $B$.  The positivity of relative entropy thus implies the positivity of $\Delta E^{grav}_B - \Delta S^{grav}_B$ for every ball-shaped region $B$ in every Lorentz frame, while monotonicity implies that it must increase if the size of the ball is increased. If one of these constraints fails to hold for some spacetime $M$ this spacetime cannot be related to any consistent state of a holographic CFT. In other words, it is unphysical.

Though (\ref{REdict}) already allows us to write down these constraints explicitly and check them for any geometry, understanding the nature of these constraints in general is difficult in the present form, with relative entropy expressed as a difference of boundary terms on $B$ and $\tilde{B}$. In \cite{lashkari2014gravitational,faulkner2014gravitation}, it was shown that to leading order in perturbations away from pure $AdS$, the set of nonlocal constraints can be recast as local constraints on the metric, and that these local constraints are precisely Einstein's equations linearized about AdS. We will now see that very similar technology can be used to rewrite the relative entropy constraints more generally, allowing a more straightforward interpretation of their implications.

\subsection{A fundamental identity}

To proceed, we will make use of a fundamental gravitational identity described recently by Hollands and Wald \cite{hollands2013stability}. Consider a one-parameter family of metrics $g_{ab}(\lambda)$, and an arbitrary vector field $X^a$. Consider also a general gravitational Lagrangian $L$ (not necessarily the actual Lagrangian for our physical system). Then the identity takes the form
\be
\label{fundI}
\boldsymbol{\omega}_L(g,dg/d\lambda,{\cal L}_X g) + {\bf \hat{E}}_L(g,dg/d\lambda) = d \boldsymbol{ \chi}_L(g,dg/d\lambda)
\ee
where $\boldsymbol{ \omega}_L(g,h_1,h_2)$ is a $d$-form whose integral over a Cauchy surface defines a natural symplectic form on the space of perturbations to a metric for the theory with Lagrangian $L$, ${\bf \hat{E}}_L$ is a d-form that vanishes if the equations of motion associated with $L$ are satisfied for $g(\lambda)$, and $\boldsymbol{ \chi}_L$ is a $(d-1)-$form whose integral over the boundary regions $B$ and the associated bulk extremal surface $\tilde{B}$ can be related respectively to a gravitational energy $\Delta E$ and entropy $\Delta S$ associated with $L$. This identity will allow us to rewrite the $(d-1)$-dimensional integrals on $B$ and $\tilde{B}$ defining $\Delta E^{grav} - \Delta S^{grav}$ in terms of a $d$-dimensional bulk integral on a bulk spacelike surface $\Sigma$ bounded by $B - \tilde{B}$.

To define the quantities appearing in the fundamental identity, consider the Lagrangian $L$ expressed as a $(d+1)$-form,
\[
\bf{L}=\mathcal{L} \boldsymbol{ \epsilon} \; .
\]
where $\boldsymbol{ \epsilon}$ is the volume form
\[
\boldsymbol{ \epsilon} =  {1 \over (d+1)!} \sqrt{-g} \epsilon_{a_1 \cdots a_{d+1}} dx^{a_1}\wedge \dots \wedge dx^{a_{d+1}} \; .
\]
For later use, we also define the lower-dimensional forms
\[
\boldsymbol{ \epsilon}_{c_1 \dots c_k} =  {1 \over (d-k+1)!} \sqrt{-g} \epsilon_{c_1 \dots c_k a_{k+1} \cdots a_{d+1}} dx^{a_{k+1}}\wedge \dots \wedge dx^{a_{d+1}} \; .
\]

Under a variation of fields this Lagrangian form varies as
\bea
\delta {\bf L} = -{\bf E}^g \cdot \delta g \boldsymbol{ \epsilon}  + d \boldsymbol{ \theta}(g,\delta g) \; ,
\eea
where ${\bf E^g}=0$ give the equations of motion for the fields, and $\theta$ is a boundary term typically called the symplectic potential current form. In this expression, $g$ is taken to represent both the metric and any other fields appearing in the Lagrangian $L$.

The term involving $\boldsymbol{ \theta}$ is the total derivative term that is produced by integration by parts when deriving the action.

The form $\boldsymbol{ \omega}$ in (\ref{fundI}) is defined in terms of $\boldsymbol{\theta}$ by
\bea
\boldsymbol{\omega}(g;\p_{\l_1}g,\p_{\l_2}g)=\p_{\l_1}\boldsymbol{ \theta}(g;\p_{\l_2}g)-\p_{\l_2}\boldsymbol{ \theta}(g;\p_{\l_1}g).
\eea
This ``symplectic current form'' plays an important role in the covariant phase space formulation of the theory. If restricted to on-shell perturbations, it is a closed and non-degenerate, and is used to define a natural symplectic form on the space of perturbations around a classical solution $g$,\footnote{As described in \cite{hollands2013stability}, it is possible to introduce canonically conjugate variables so that the symplectic form becomes simply
\[
W_\Sigma(g;\delta_1 g, \delta_2 g) = - {1 \over 16 \pi} \int \sqrt{h}[ \delta_1 h_{ab} \delta_2 p^{ab} - \delta_1 h_{ab} \delta_2 p^{ab} ]\; .
\]}
\be
\label{defW}
W_\Sigma(g,\gamma_1, \gamma_2)  = \int_\Sigma \boldsymbol{ \omega}(g, \gamma_1, \gamma_2).
\ee
To define the form $\boldsymbol{ \chi}$ appearing in (\ref{fundI}), we consider the Noether current associated to diffeomorphisms generated by the vector field $X$. Expressed as a differential form, this is
\[
{\bf J}_X = \boldsymbol{ \theta}(g, {\cal L}_X g) - i_X {\bf L}(g) \; .
\]
Current conservation implies that this form can be expressed as a total derivative plus a term that vanishes when the equations of motion are satisfied,
\[
{\bf J}_X = d {\bf Q}_X + {\bf C}_X \; ,
\]
In terms of these quantities, we have
\bea\label{chiform}
\boldsymbol{ \chi}(g,{ d  \over  d \lambda} g) = {d \over d \lambda} {\bf Q}_X(g) - i_X \boldsymbol{\theta} (g ; {d \over d \lambda} g ) \; .
\eea
Finally, the term in (\ref{fundI}) involving the equations of motion is defined to be
\[
{\bf \hat{E}_L}(g,{d \over d\lambda} g) = i_X({\bf E}(g) \cdot {d \over d \lambda} g)- {d \over d \lambda} {\bf C}_X(g)
\]
All of these quantities depend on which gravitational Lagrangian we choose. For the case of pure Einstein gravity with a cosmological constant, we have \cite{hollands2013stability, faulkner2014gravitation}
\bea
\label{Puredefs}
\mathcal{L} &=& {1 \over 16 \pi} R - \Lambda  \cr
E^g_{ab} &=& {1 \over 16 \pi} (R_{ab} - {1 \over 2} g_{ab} R) + {1 \over 2} g_{ab} \Lambda \cr
{\bf C}_a &=& 2 X^a E^g_{ab} \epsilon^b \cr
{\bf Q}_X &=& {1 \over 16 \pi} \nabla^a X^b \epsilon_{ab} \cr
\boldsymbol{\theta} &=& {1 \over 6 \pi} \epsilon_a (g^{ac} g^{bd}-g^{ad} g^{bc}) \nabla_d {d \over d \lambda} g_{bc} \cr
\boldsymbol{\omega} &=& {1 \over 16 \pi} \epsilon_a P^{abcdef} (\gamma^2_{bc} \nabla_d \gamma^1_{ef} - \gamma^1_{bc} \nabla_d \gamma^2_{ef}) \cr
P^{abcdef} &=& g^{ae} g^{fb} g^{cd} - {1 \over 2} g^{ad} g^{be} g^{fc} - {1 \over 2} g^{ab} g^{cd} g^{ef} - {1 \over 2} g^{bc} g^{ae} g^{fd} + {1 \over 2} g^{bc} g^{ad} g^{ef}
\eea
Using (\ref{chiform}) and the equations above, we find that
\be
\label{defchi}
\chi(\gamma,X) = {1 \over 16 \pi} \epsilon_{ab} \left\{\gamma^{ac} \nabla_c X^b - {1 \over 2} \gamma_c{}^c \nabla^a X^b + \nabla^b \gamma^a {}_c X^c - \nabla_c \gamma^{ac} X^b + \nabla^a \gamma^c {}_c X^b \right\} \; .
\ee

\subsection{Bulk integral for relative entropy}

We now consider a one-parameter family of asymptotically AdS spacetimes $M(\lambda)$, and the family of extremal surfaces $\tilde{B}(\lambda)$ associated with some fixed ball-shaped boundary region $B$. In \cite{hollands2013stability}, it was shown that it is always possible to choose metrics $g(\lambda)$ such that the extremal surface $\tilde{B}(\lambda)$ has a fixed coordinate location, and such that the Killing vector $\xi_B^a$ defined in section 1 continues to satisfy
\be
\label{Kcond}
(\xi_B)|_{\tilde{B}} = (\nabla_{a} (\xi_B)_{b} + \nabla_{b} (\xi_B)_{a})|_{\tilde{B}}  = 0 \; .
\ee
That is, $\xi$ continues to behave as a Killing vector near the extremal surface $\tilde{B}$.

Consider the gravitational expression for relative entropy evaluated for this family of spacetimes,
\[
S(g(\lambda)||g_0) \equiv \Delta E^{grav}(g(\lambda)) - \Delta S^{grav}(g(\lambda)) \; .
\]
Using the fundamental identity (\ref{fundI}) we now show that the first derivative $ \partial_\lambda S(g(\lambda)||g_0)$ can be written as an integral over a spacelike surface $\Sigma$ bounded by $B$ and $\tilde{B}$.

First, we note that
\be
\label{intB}
\int_{B} \boldsymbol{ \chi}(g, dg/d\lambda) = {d \over d \lambda} E^{grav}_B \; .
\ee
This was argued in general in \cite{faulkner2014gravitation}.

Next, we have that
\[
\int_{\tilde{B}} \boldsymbol{ \chi}(g, dg/d\lambda) = {d \over d \lambda} S^{grav}_B \; .
\]
This follows by the vanishing of $\xi$ on $\tilde{B}$, which gives
\[
\boldsymbol{ \chi}|_{\tilde{B}} = {\bf Q}_\xi|_{\tilde{B}} \; ,
\]
and the result
\be
\label{intBt}
\int_B {\bf Q}_{\xi}  = {1 \over 4} A \; .
\ee
This holds in the unperturbed spacetime since $Q$ is the Noether charge associated with the Killing vector $\xi_B$, which defines the Wald entropy of the bifurcate Killing horizon $\tilde{B}$. As shown in \cite{hollands2013stability}, this continues to hold in the perturbed spacetime because of the gauge condition (\ref{Kcond}).

Combining these results, we have that
\bea
{d \over d \lambda} S(g(\lambda)||g_0)  &=& {d \over d \lambda} (E^{grav}(g(\lambda)) - S^{grav}(g(\lambda))) \cr
&=& \int_{\tilde{B}} \boldsymbol{ \chi} - \int_B \boldsymbol{ \chi}  \cr
&=& \int_{\partial \Sigma} \boldsymbol{ \chi} \cr
&=& \int_\Sigma d \boldsymbol{ \chi}
\eea
Finally, using the identity (\ref{fundI}), we obtain
\be
\label{dlambda}
{d \over d \lambda} S(g(\lambda)||g_0) = W_\Sigma(g ; {d \over d \lambda} g, {\cal L}_\xi g) +\int_\Sigma \left\{i_X(E(g)\cdot {d \over d \lambda} g) - {d \over d \lambda} C_X(g) \right\}
\ee
where the last line makes use of the identity (\ref{fundI}).

This is the fundamental relation that we will make use of below when translating constraints on relative entropy to constraints on geometry. Primarily, we will make use of this identity for the case where the Lagrangian is chosen to be that for pure Einstein gravity with cosmological constant, so that all quantities in the expression above are purely gravitational quantities. However, we can alternatively choose to consider the case where the various quantities are defined with respect to the Lagrangian for Einstein gravity coupled to matter. In this case, assuming that curvature tensors do not appear in the matter part of the Lagrangian, the results (\ref{intB}) and (\ref{intBt}) remain valid, so the expression (\ref{dlambda}) is also correct when $W$, $E$, and $C$ are constructed starting from the full Lagrangian including matter. In this case, the terms involving $E(g)$ and $C_X(g)$ vanish on shell, since these are built from the tensors appearing in the full equations of motion. Thus, we have that
\be
\label{mattfromgrav}
 W^{full}_\Sigma(g ; {d \over d \lambda} g, {\cal L}_\xi g) =  W_\Sigma(g ; {d \over d \lambda} g, {\cal L}_\xi g) +\int_\Sigma \left\{i_X(E(g)\cdot {d \over d \lambda} g) - {d \over d \lambda} C_X(g) \right\}
\ee
where the expressions on the right are purely gravitational.

\subsection{First order results}

We first consider the result (\ref{dlambda}) evaluated at $\lambda=0$. Since $\xi$ is a Killing vector of the unperturbed metric, we have ${\cal L}_\xi g = 0$. so the term $W_\Sigma(g ; {d \over d \lambda} g, {\cal L}_\xi g)$ vanishes. Also, the unperturbed AdS metric satisfies the vacuum Einstein equations, so the term $i_X(E(g)\cdot {d \over d \lambda} g)$ also vanishes. Thus, using (\ref{Puredefs}), we have
\be
\label{REfirst}
{d \over d \lambda} S(g(\lambda)||g_0)|_{\lambda = 0}  = -\int_\Sigma  {d \over d \lambda} C_X(g) = -2 \int_\Sigma \xi^a {d E^g_{ab} \over d \lambda}  \epsilon^b
\ee
where $E^g_{ab}$ are the gravitational equations. Positivity of relative entropy in the CFT implies that the relative entropy is minimized for the vacuum state, so the first order variation must vanish. Gravitationally, this implies that the left side of (\ref{REfirst}) must vanish, so we have that
\[
\int_\Sigma \xi^a {d E^g_{ab} \over d \lambda}  \epsilon^b = 0 \; .
\]
As shown in \cite{faulkner2014gravitation}, if this holds for all regions $\Sigma$ associated with any ball $B$ in any Lorentz frame, we must have that $d E^g_{ab}/ d \lambda = 0$, that is, the metric $g(\lambda)$ must satisfy the Einstein equation to first order in $\lambda$. Thus, for spacetimes $M(\lambda)$ which geometrically encode the entanglement entropies of CFT states via the HRT formula, the constraints of relative entropy positivity at first order in $\lambda$ are precisely the linearized gravitational equations.

\subsection{Second order results: the gravity dual of Fisher Information}

Next, consider the $\lambda$ derivative of the result (\ref{dlambda}) evaluated at $\lambda = 0$. Defining $\gamma = dg/d \lambda|_{\lambda=0}$ as the first order metric perturbation we find (using ${\cal L}_\xi g = E(g) = d /d \lambda (E(g)) = 0$)
\be
\label{REsecond}
{d^2 \over d \lambda^2} S(g(\lambda)||g_0)|_{\lambda = 0}  = W_\Sigma(g, \gamma, {\cal L}_\xi \gamma) -2 \int_\Sigma \xi^a {d^2 E^g_{ab} \over d \lambda^2}  \epsilon^b
\ee
Consider first the case where we have a holographic CFT dual to some known theory of Einstein gravity coupled to matter, and where the quantities $E^g$ and $W$ are defined with respect to the full Lagrangian.
Then $E^g$ represent the full equations of motion for the theory, which should vanish for
the one-parameter family of field configurations $g(\lambda)$ dual to holographic CFT states $|\Psi(\lambda) \rangle$. Thus, we have simply:
\be
\label{REsecond2}
{d^2 \over d \lambda^2} S(g(\lambda)||g_0)|_{\lambda = 0}  = W_\Sigma(g, \gamma, {\cal L}_\xi \gamma)
\ee
The left side is precisely the gravitational dual of the Fisher Information $\langle \delta \rho , \delta \rho \rangle$, while the right side was defined in \cite{hollands2013stability} as the {\it canonical energy} \be
\label{canonical2}
{\cal E}(\delta g, \delta g) \equiv  W_\Sigma(g, \gamma, {\cal L}_\xi \gamma) \; .
\ee
Thus, for holographic CFTs in the classical limit, we have that {\it Fisher Information is dual to canonical energy},
\[
\langle \delta \rho_B , \delta \rho_B \rangle = {\cal E}_B(\delta g, \delta g) \; .
\]
More generally, we can promote ${\cal E}_B$ to a bilinear form on perturbations,
\[
{\cal E}_B(\delta g_1, \delta g_2) \equiv W_\Sigma(g, \delta g_1, {\cal L}_\xi \delta g_2) \; ,
\]
which can be shown to be symmetric. This quantity is dual to the Fisher Information metric defined above,
\[
\langle (\delta \rho_B)_1 , (\delta \rho_B)_2 \rangle = {\cal E}_B(\delta g_1, \delta g_2) \; .
\]
Since the Fisher information and the Fisher information metric must be non-negative, it must be that the corresponding gravitational quantities are also non-negative. Thus, the positivity of relative entropy at second order implies the positivity of canonical energy. Specifically, for any one parameter family $g(\lambda)$ of physical asymptotically AdS spacetimes, and for any ball-shaped region $B$ on the boundary, we must have ${\cal E}_B(\delta g, \delta g) > 0$. It should be possible to demonstrate this directly in specific consistent classical theories of gravity.

\subsection{Gravitational constraints from positivity of Fisher Information}

In general, the expression for canonical energy defined in the previous section depends on both the metric and the matter fields for the theory. However, using the result (\ref{mattfromgrav}), it is always possible to rewrite it (using the equations of motion) as a purely gravitational expression.  Defining the gravitational part of canonical energy
\[
{\cal E}^{grav}(\gamma, \gamma) = W^{grav}_\Sigma(g, \gamma, {\cal L}_\xi \gamma)  \; ,
\]
we find from (\ref{mattfromgrav}) that
\be
\label{metricConstr}
{\cal E}(\gamma, \gamma) = {\cal E}_B^{grav}(\gamma, \gamma) - 2 \int_\Sigma \xi^a {d^2 E^g_{ab} \over d \lambda^2}  \epsilon^b \ge 0
\ee
This gives a purely geometrical constraint on asymptotically AdS spacetimes that can arise in consistent theories for which the HRT formula holds (expected to be theories with Einstein gravity coupled to matter in the classical limit).

Note that ${\cal E}_B$ is calculated using only the first order perturbation $\gamma = dg/d \lambda|_{\lambda=0}$, which must solve the Einstein equations linearized about AdS. The second term involves also the metric at second order. Thus, we can think of the relation (\ref{metricConstr}) as constraining the ${\cal O}(\lambda^2)$ terms in the metric in terms of the ${\cal O}(\lambda)$ terms. We provide some explicit examples in section 4 below.

Another useful form of the constraint is obtained from the expression (\ref{metricConstr}) by making use of the general expression for the gravitational equations (recalling the normalization of $E^g_{ab}$ in (\ref{Puredefs}))
\be
\label{EE}
E^g_{ab} = {1 \over 2} T^{matt}_{ab} \; .
\ee
This gives
\be
\label{metricConstr1}
{\cal E}(\gamma, \gamma) = {\cal E}_B^{grav}(\gamma, \gamma) -  \int_\Sigma \xi^a T^{(2)}_{ab} \epsilon^b \ge 0
\ee
where $T^{(2)}_{ab}$ are the terms in the matter stress tensor at second order in $\lambda$. Thus, the positivity of Fisher information provides a constraint the behavior of the matter stress-energy tensor that should hold in any consistent theory.

As a more explicit example, if we use Fefferman-Graham coordinates $ds^2 = (dz^2 + dx_\mu dx^\mu)/z^2$, and consider the ball $B = \{t=0, |\vec{x}| \le R\}$, the Killing vector $\xi_B$ is
\be
\label{defxi}
\xi_B  = - \frac{2\pi}{ R}  (t-t_0) [z \p_z + (x^i-x^i_0) \p_i ] + \frac{\pi}{R} [R^2 - z^2 - (t-t_0)^2 - (\vec{x}-\vec{x_0})^2] \, \p_t
\ee
so the constraint (\ref{metricConstr1}) is
\be
\label{Constrexplicit}
\int_{x^2 + z^2 < R^2} dz d^{d-1} x {\pi(R^2 - z^2 - \vec{x}^2) \over R z^{d-1}} T^{(2)}_{00}  \ge - {\cal E}_B^{grav}(\gamma, \gamma) \; .
\ee
Since the gravitational contribution to canonical energy must be positive on its own, we see that positivity of the matter stress-tensor (more generally, the weak energy condition) will guarantee that the relative entropy constraint is satisfied. However, it is also possible to satisfy (\ref{Constrexplicit}) with a certain amount of negative energy. Thus, the positivity of relative entropy implies a somewhat weaker integrated energy condition, as pointed out for special cases in \cite{Lin:2014hva,Lashkari:2014kda}. We give some more explicit examples derived from this constraint in section 4 below.

Finally, we note that equation (84) in \cite{hollands2013stability} gives an illuminating expression for the gravitational part of canonical energy,
\[
{\cal E}_B^{grav}(\gamma, \gamma) = -\int_\Sigma \xi^a T^{grav (2)}_{ab} \epsilon^b - \int_B {d^2 \over d \lambda^2} Q_{\xi}(g + \lambda \gamma)
\]
where
\[
T^{grav (2)}_{ab} = -{d^2 \over d \lambda^2} E^g_{ab}(g + \lambda \gamma)|_{\lambda=0}
\]
is the expression quadratic in the first order metric perturbation that provides the source term in the equation determining the second order perturbation when solving Einstein's equations perturbatively. Thus, we have
\[
{\cal E}_B(\gamma, \gamma) = -\int_\Sigma \xi^a (T^{(2)}_{ab} + T^{grav (2)}_{ab}) \epsilon^b + {\rm boundary \; term} \; .
\]
Up to the boundary term, this is exactly the ``Rindler energy'' associated with the Killing vector $\xi_B$ in the wedge $R_B$, including perturbative contributions from both the metric perturbation and the matter fields. Thus, it is indeed the ``canonical'' expression for energy computed with respect to the timelike Killing vector $\xi$ in the background geometry.

\section{Examples}

In this section, we provide some examples to illustrate the calculation of canonical energy for perturbations to asymptotically AdS spacetimes. Such calculations are necessary to provide a more explicit form of the energy condition (\ref{metricConstr}), to check that the condition is satisfied for particular cases, or to prove that this condition is satisfied in general for a specific theory (e.g. pure gravity).

\subsection{Transformation to Hollands-Wald gauge}

The main challenge for calculations is that the results of section 3 (and of \cite{hollands2013stability}) make use of the assumed gauge choice that the extremal surfaces $\tilde{B}$ for the family of spacetimes $g(\lambda)$ all have the same coordinate description and that the Killing vector $\xi_B$ of the unperturbed spacetime continues to satisfy (\ref{Kcond}). It will be useful to have a procedure that allows us to calculate the canonical energy for a perturbation given in some more general gauge.

Thus, suppose that $g$ is some background satisfying the equations of motion, $h$ is some perturbation satisfying the linearized equations about the background $g$ (but not necessarily the gauge condition), and $K$ is the Killing vector in the unperturbed space. Then there is some metric perturbation $\gamma$ satisfying the gauge condition that is related to $h$ by a gauge transformation,
\[
\gamma = h + {\cal L}_V g
\]

To determine the required gauge transformation $V$, we begin with the condition that the original extremal surface remains extremal under the perturbation $\gamma$. To derive an explicit condition on $V$, it is convenient to choose coordinates for the unperturbed spacetime such that the extremal surface is described by
\[
X^i = \sigma^i \qquad X^A = X_0^A
\]
where $\sigma^i$ are the coordinates that we use to parametrize the surface, and $X_0^A$ are constants. Then our condition is that for the area functional $A(X + \delta X,g + \gamma)$, the term at order $\delta X$ vanishes both for $\gamma=0$ and at linear order in $\gamma$. In calculating this term, we can use the simplification that all derivatives of $X^A(\sigma)$ vanish. The final result is
\be
\label{extrcond_general}
(\nabla_i \gamma^i {}_A - {1 \over 2} \nabla_A \gamma^i {}_i)_{\tilde{B}} = 0
\ee
where $i$ runs over the directions along the surface $\tilde{B}$ and $A$ runs over the transverse directions. We obtain a condition on $V$ by the substitution
\be
\label{gammafromh}
\gamma_{ab} = (h + {\cal L}_V g)_{ab} = h_{ab} + \nabla_a V_b + \nabla_b V_a \; .
\ee

Explicitly, this gives
\be
\label{EcondV}
(\nabla_i \nabla^i V_A + [\nabla_i, \nabla_A] V^i + \nabla_i h^i {}_A - {1 \over 2} \nabla_A h^i {}_i )_{\tilde{B}}= 0 \; .
\ee

The condition that $K$ continues to satisfy ${\cal L}_K g_{\tilde{B}} = \nabla_{(a} K_{b)}|_{\tilde{B}}=0$ in the perturbed geometry gives
\[
{\cal L}_K h |_{\tilde{B}} = 0 \; ,
\]
or explicitly,
\[
(\gamma^c {}_{b} \nabla_a K_c + \gamma^c {}_{a} \nabla_b K_c)_{\tilde{B}} = 0 \; .
\]
Since $\nabla_{(a} K_{b)} = 0$ and $\nabla_{[a} K_{b]} \propto \epsilon_{a b}$, this is equivalent to
\be
\label{Kcond_general}
(\gamma^c {}_b \epsilon_{ac} + \gamma^c {}_a \epsilon_{bc})_{\tilde{B}} = 0
\ee
where $\epsilon_{ab} = n_1^a n_2^b - n_2^a n_1^b$ is the binormal to the surface $\tilde{B}$. Taking the various components of this expression in the normal and tangential directions, we find
\be
\label{Kcond_comp}
(\gamma^i {}_A)_{\tilde{B}} = 0 \qquad \qquad (\gamma^A {}_{D} - {1 \over 2} \delta^A {}_{D}  \gamma^C {}_C)_{\tilde{B}} = 0 \; .
\ee
Finally, using (\ref{gammafromh}), we find that the conditions on $V$ are
\bea
\label{KcondV}
(h_{i A} + \nabla_i V_A + \nabla_A V_i)_{\tilde{B}} &=& 0 \cr
(h^A {}_{D} - {1 \over 2} \delta^A {}_{D}  h^C {}_C + \nabla^A V_D + \nabla_D V^A - \delta^A {}_D  \nabla_C V^C)_{\tilde{B}} &=& 0 \; .
\eea

To summarize, given a metric perturbation $h$, the equations (\ref{EcondV}) and (\ref{KcondV}) determine the conditions on $V$ so that the gauge transformation gives the metric perturbation $\gamma$ equivalent to $h$ but satisfying the gauge conditions.

\subsection{Calculating canonical energy from $h$ and $V$}

The canonical energy is calculated using the definition (\ref{canonical2}) together with (\ref{defW}) and (\ref{Puredefs}), where the metric perturbation $\gamma$ is assumed to obey the gauge constraint. Using the results of the previous section, we can write $\gamma$ for some arbitrary metric perturbation using (\ref{gammafromh}), where $V$ is required to satisfy the conditions equations (\ref{EcondV}) and (\ref{KcondV}) at the surface $\tilde{B}$. We will now see that the canonical energy can be evaluated using the same expression as in (\ref{canonical2}), applied to $h$, plus a boundary integral that depends on $h$ and $V$.

To begin, we note that
\bea
\omega(g,\gamma,{\cal L}_K \gamma)
&=& \omega(g,h + {\cal L}_V g,{\cal L}_K (h + {\cal L}_V g)) \cr
&=& \omega(g,h,{\cal L}_K h) + \omega(g,h + {\cal L}_V g,{\cal L}_{[K,V]} g) - \omega(g,{\cal L}_K h,  {\cal L}_V g)
\label{omega}
\eea
where we have used that ${\cal L}_K g = 0$ and
\[
{\cal L}_K {\cal L}_V g = [{\cal L}_K, {\cal L}_V] g = {\cal L}_{[K,V]} g \; .
\]
In the final expression, the commutator of vector fields is defined as
\[
[K,V]^a = K^b \partial_b V^a - V^b \partial_b K^a \; .
\]
Using the fundamental identity (\ref{fundI}), we have
\be
\label{dchi}
\omega(g, \gamma,{\cal L}_\xi g) = d \chi(\gamma,X)
\ee
for any $g$ and $\gamma$ satisfying the equations of motion, where $\chi$ is given in (\ref{defchi}).

The second and third terms in (\ref{omega}) take the form of the left side of (\ref{dchi}), so all can be written as derivatives of a form. Thus, we can write
\be
\label{defomega}
\omega(g,\gamma,{\cal L}_K \gamma) = \omega(g,h,{\cal L}_K h) + d \rho
\ee
where
\be
\label{defrho}
\rho = \chi(h + {\cal L}_{V} g,[K,V])  - \chi({\cal L}_K h,V)  \; .
\ee

In the integral (\ref{defW}) defining canonical energy, the integral over $d \rho$ can be converted to a boundary integral (over $\partial \Sigma = \tilde{B} - B$) using Stokes' theorem. Since the conditions on $V$ are localized to $\tilde{B}$, we can always choose $V$ to vanish at the other boundary so that $\int_B \rho = 0$. In this case, we have
\be
\label{calcE}
{\cal E}_B = \int_\Sigma \omega(g, h, {\cal L}_K h) + \int_{\tilde{B}} \rho(h,V) \; .
\ee
Thus, given a metric perturbation $h$ in some general gauge, we can compute the canonical energy for the region associated with a ball $B$ by finding $V$ satisfying the conditions (\ref{EcondV}) and (\ref{KcondV}) and vanishing near $B$ and evaluating (\ref{calcE}). Note that we don't need the explicit form of $V$ everywhere; rather, we need only determine $V$ (and some of its derivatives) at the surface $\tilde{B}$.

\subsection{Example: perturbations to Poincare $AdS_3$}

We now consider the specific example of perturbations to $AdS_3$. It will be convenient to use polar coordinates for the unperturbed metric
\[
ds^2 = {1 \over r^2 \cos^2 \theta}(-dt^2 + dr^2 + r^2 d \theta^2) \; ,
\]
so that the extremal surface for a region $B = {x \in [-R,R], t =0}$ is given as
\be
\label{AdS_extr}
\tilde{B} = \{t=0 ,  r=R \}
\ee
with $\theta$ chosen as the embedding coordinate. In these coordinates, the Killing vector $K=\xi_B$ in the unperturbed geometry is
\[
\xi_B = -{\pi \over R}(-R^2+t^2+r^2) \partial_t -{2 \pi \over R} r t \partial_r \; .
\]

For perturbations to the background, the condition (\ref{AdS_extr}) for the surface to remain extremal become
\bea
\label{extrcondAds3}
[\cos \theta \; (\partial_\theta \gamma_{r \theta} - {1 \over 2} \partial_r \gamma_{\theta \theta}) - \sin \theta \; \gamma_{r \theta}]|_{\tilde{B}} &=& 0 \cr
[\cos \theta \; (\partial_\theta \gamma_{t \theta} - {1 \over 2} \partial_t \gamma_{\theta \theta}) - \sin \theta \; \gamma_{t \theta}]_{\tilde{B}} &=& 0 \; .
\eea
while the condition for $\xi$ to satisfy the Killing vector condition on $\tilde{B}$ are
\be
\label{KcondAdS}
\gamma_{tr}|_{\tilde{B}} = \gamma_{t \theta}|_{\tilde{B}} = \gamma_{r \theta}|_{\tilde{B}} = (\gamma_{tt} + \gamma_{rr})|_{\tilde{B}} = 0
\ee
Translating these to the explicit conditions (\ref{KcondV}) and (\ref{EcondV}) on $V$ give
\bea
\partial_\theta^2 V_t - 3 \tan(\theta) \partial_\theta V_t - 2 V_t &=& \tan(\theta) h_{t\theta} + {1 \over 2} \partial_t h_{\theta\theta} - \partial_\theta h_{t\theta} \cr
\partial_\theta^2 V_r - 3 \tan(\theta) \partial_\theta V_r - 2 V_r &=& \tan(\theta) h_{r\theta} + {1 \over 2} \partial_r h_{\theta\theta} - \partial_\theta h_{r\theta} \cr
\partial_t V_r + \partial_r V_t + {2 \over r} V_t &=& -  h_{tr} \cr
\partial_t V_t + \partial_r V_r + {2 \over r} V_r &=& - {1 \over 2} (h_{tt} + h_{rr}) \cr
\partial_\theta V_t + \partial_t V_\theta - 2 \tan(\theta) V_t &=& - h_{t \theta} \cr
\partial_\theta V_r + \partial_r V_\theta - 2 \tan(\theta) V_r &=& - h_{r \theta}
\label{VcondAdS3}
\eea
All of these equations are required to hold on the surface $r=R$.  Given a perturbation $h_{ab}$ we must then use these equations to determine $V$ and its derivatives on this surface, which allows us to calculate the canonical energy for this perturbation using (\ref{calcE}).

\subsubsection*{Homogeneous perturbations}

As an example, we consider a perturbation to the planar black hole geometry. In Fefferman-Graham coordinates, this geometry is described by
\be
\label{BTZ}
ds^2 = {1 \over z^2}(dz^2 + (1 + \mu z^2/2)^2 dx^2 - (1 - \mu z^2/2)^2 dt^2) \; .
\ee
In the polar coordinates that we are using, the perturbation to first order in $\mu$ is given by
\be
\label{hompert}
h_{rr} = \mu \sin^2 \theta \qquad h_{tt} = \mu \qquad h_{r \theta} = \mu r \sin \theta \cos \theta \qquad h_{\theta \theta} = \mu r^2 \cos^2 \theta \; .
\ee
To solve (\ref{VcondAdS3}), we can choose $V$ of the form
\[
V = \mu (V^r\p_r+V^\theta\p_\theta) \; .
\]
In this case, we find that the equations (\ref{VcondAdS3}) are satisfied if and only if the following conditions are satisfied at $\tilde{B}$:
\bea
\partial_\theta^2 V_r - 3 \tan \theta \partial_\theta V_r - 2 V_r - 2 r \sin^2 \theta &=& 0 \cr
\partial_r V_\theta + \partial_\theta V_r - 2 \tan \theta V_r + r \sin \theta \cos \theta &=& 0 \cr
 2 \partial_r V_r + {4 \over r} V_r + 2  - \cos^2 \theta &=& 0
\eea
These require that:
\bea
\label{Vbcs}
V_r(R, \theta)|_{r=R} &=& R \left( {1 \over 6}  (\cos^2 \theta - 2) + {C_2 \over \cos^2 \theta}  + {C_1 \sin \theta \over \cos^2 \theta } \right)\cr
\partial_r V_r (R, \theta)|_{r=R} &=& {1 \over 6} (\cos^2 \theta - 2) - 2 {C_2 \over \cos^2 \theta}  -2 {C_1 \sin \theta \over \cos^2 \theta } \cr
\partial_r V_\theta(R,\theta)|_{r=R} &=& R \left( -{1 \over 3} \cos \theta \sin \theta - {2 \over 3} {\sin \theta \over \cos \theta} - {C_1 \over \cos \theta} \right)
\eea
We choose $C_1=C_2=0$ in order that $V$ is well-behaved at the boundary (where $\cos(\theta) \to 0$). Fortunately, these are the only properties of $V$ that will be required for our calculation.
%

We are now ready to calculate the canonical energy using (\ref{calcE}). Making use of the definition (\ref{Puredefs}), we find
\[
\omega(g,h,{\cal L}_\xi h)_\Sigma = {1 \over R} \left[ -{1 \over 2} r^4 \cos^3 \theta \right] dr \wedge d \theta \; ,
\]
so that
\be
\label{wpart}
\int_\Sigma \omega(g,h,{\cal L}_\xi h)_\Sigma = \frac{1}{R}\int_0^R dr \int_{-\pi \over 2}^{\pi \over 2} d \theta \left[ -{1 \over 2} r^4 \cos^3 \theta \right]  = - {2 \over 15} R^4 \; .
\ee
Using (\ref{defrho}) and (\ref{Puredefs}), we find that
\[
\rho|_{\tilde{B}} = -{R^4 \over 12} \cos^3 (\theta) (2 \cos^2 (\theta) - 3) d \theta + \rho_r dr
\]
where $\rho_r$ depends on the specific form of $V_\theta$ but is not needed for our calculation.
This gives
\be
\label{rhopart}
\int_{\tilde{B}} \rho = {7 \over 45} R^4 \; .
\ee
Combining (\ref{wpart}) and (\ref{rhopart}) as in (\ref{calcE}) to calculate the (gravitational part of) canonical energy we find
\be
\label{Eres}
{\cal E}_{B}(\gamma,\gamma) = {R^4 \over 45} \; .
\ee
In the case of pure gravity, or where no other fields are turned on in the bulk, this is the complete result for the canonical energy associated with the wedge $R_B$ for a ball $B$ of radius $R$. The positivity of Fisher information required that this be positive, so we see that the constraints are satisfied.

\subsubsection*{Comparison with relative entropy}

As a check we now compare the result with the second derivative $E^{grav} - S^{grav}$ about pure AdS. Using the metric (\ref{BTZ}), we can compute the extremal surface $B$ for arbitrary $\mu$ and compare its area with the unperturbed result. Using calculations in \cite{Lashkari:2014kda} we have that
\[
S(\mu) - S_{vac} = {1 \over 2 G} \left[ \int_0^{z_0} {dz \over z} \left\{{1 \over \sqrt{1 - {z^2 f(z0) \over z0^2 f(z)}}}-1\right\} - \ln \left( {2R \over z_0} \right) \right]
\]
where $f(z) = (1 + \mu/2 z^2)^2$, and $z_0$ is related to $R$ by
\[
R = \int_0^{z_0} {1 \over \sqrt{{f^2(z) z_0^2 \over f(z_0) z^2} - f(z)}} \; .
\]
Working perturbatively in $\mu$, we find
\[
S(\mu) - S_{vac} = \mu {R^2 \over 6 G} - {R^4 \over 90 G} \mu^2 + {\cal O} (\mu^3) \; .
\]

To find $\Delta E$, we use that
\[
\langle T_{\mu \nu} \rangle = {1 \over 8 \pi G} h_{\mu \nu}^0
\]
and
\[
\Delta E = 2 \pi \int_{-R}^R {R^2 - x^2 \over 2 R} \langle T_{tt} \rangle \; .
\]
Combining these and using that $h_{tt}^{(0)} = \mu$, we get
\[
\Delta E = \mu {R^2 \over 6 G} \; .
\]
Thus, to second order in $\mu$, we find that the relative entropy is
\[
\Delta E - \Delta S = \mu^2 {R^4 \over 90 G} \; ,
\]
so (setting $G$ to 1),
\[
{d^2 \over d \mu^2}\left(\Delta E - \Delta S  \right) = {R^4 \over 45} \; .
\]
This agrees precisely with our expression above.

\subsubsection*{Constraints for theories with matter}

The result (\ref{Eres}) gives the canonical energy ${\cal E}_B$ associated with the homogeneous first order perturbation (\ref{hompert}) in the case where the metric is the only field turned on in the bulk. Since we expect that the geometry (\ref{BTZ}) corresponds (for positive $\mu$) to a physically consistent state (the thermal state of a holographic CFT), the positivity of canonical energy was fully expected; our calculation serves as a consistency check for the HRT formula.

More generally, consider a theory with Einstein gravity coupled to matter. First order perturbations to pure AdS are still governed by the linearized Einstein equations, since the matter stress tensor typically has only quadratic and higher order terms in the fields. Thus, the perturbation (\ref{hompert}) still represents a consistent deformation in this case. However at second and higher order, the metric can differ from (\ref{BTZ}) in the case when matter fields are present. In this case, the full expression (\ref{metricConstr}) for canonical energy includes contributions from the second order terms in the metric, or equivalently, via (\ref{metricConstr1}), from the matter stress-energy tensor. In the latter form, equation (\ref{Constrexplicit}) together with (\ref{Eres}) give that the positivity constraint is :
\be
\label{Constrexplicit2}
\int_{x^2 + z^2 < R^2} dz dx {\pi(R^2 - z^2 - x^2) \over R z} T^{(2)}_{00}  \ge - {R^4 \over 45}\; .
\ee
To express this directly as a constraint on the geometry in the case of a static, translation-invariant spacetime, we write the metric $g(\mu)$ as
\[
ds^2 = ds^2_{AdS} + \mu (dx^2 + dt^2) + \mu^2 (h^{(2)}_{tt}(z)dt^2+h^{(2)}_{xx}(z)dx^2) + {\cal O}(\mu^3) \; .
\]
Then after integrating over $x$ and integrating by parts to eliminate $z$ derivatives on $h$ (assuming $h$ vanishes at the $z=0$), (\ref{metricConstr}) gives
\be
\label{conexp}
\int_0^R {z^3 h^{(2)}_{xx}(z) \over \sqrt{R^2-z^2}} \le {8 R^5 \over 45} \; .
\ee
This constraint must hold for all possible $R$. As a special case, we can consider this constraint in the limit of small $R$ to place constraints on the coefficients of $h^{(2)}_{xx}$ expanded as a power series in $z$. We have checked that this precisely reproduces the constraints from positivity of relative entropy obtained in \cite{Lashkari:2014kda}.

As discussed in \cite{Lashkari:2014kda}, for the case of homogeneous perturbations to $AdS_3$, it is possible to come up with stronger constraints by demanding positivity of relative entropy with the reference state chosen to be the thermal state $\rho_T$ with the same energy expectation value as the state $|\Psi \rangle$. For such a thermal state, the modular Hamiltonian for an interval is an integral over the region of an expression linear in components of the stress-energy tensor. By construction, the stress-energy tensor expectation values match for $|\Psi \rangle$ and $\rho_T$, so $\Delta E_B$ in (\ref{RE1}) vanishes, and the second order constraint of relative entropy positivity becomes $\partial_\lambda^2 (S_A(\rho_T) - S_A(|\Psi \rangle)) \ge 0$. Now, let $g(\lambda)$ and $g_T(\lambda)$ be metrics describing the spacetimes dual to $|\Psi \rangle$ and $\rho_T$. Taking the difference of the equation (\ref{REsecond}) applies to the two states, we find that
\[
\partial_\lambda^2 (S_A(\rho_T) - S_A(|\Psi \rangle)) = -2 \int_\Sigma \xi^a {d^2 E^g_{ab} \over d \lambda^2} \epsilon^b \; ,
\]
since the first order perturbations $\gamma$ and the $\Delta E^{grav}$ depend only on the boundary stress tensor and are thus the same for both solutions. Therefore, rewriting the Einstein tensor here in terms of the matter stress tensor using (\ref{EE}), we have that the positivity constraint is precisely
\[
-\int_\Sigma \xi^a  T^{(2)}_{ab}  \epsilon^b \ge 0 \; ,
\]
that is, the Rindler energy computed from the second order matter stress-energy tensor must be positive for each Rindler wedge. For the example of a spatial interval, the explicit constraint (\ref{conexp}) on the second order metric is strengthened to
\be
\label{conexp}
\int_0^R {z^3 h^{(2)}_{xx}(z) \over \sqrt{R^2-z^2}} \le {2 R^5 \over 15} \; .
\ee

\section{Discussion}

In this paper, we have shown the canonical energy for perturbations to Rindler wedges of pure $AdS$ spacetime may be identified with the quantum Fisher information which compares the density matrix for the corresponding boundary region with the vacuum density matrix for the same region. Conversely, for any CFT states $|\Psi(\lambda) \rangle$ whose entanglement entropies are encoded holographically in dual spacetimes $M(\lambda)$ via the covariant holographic entanglement entropy formula, the Fisher information of a ball $B$ must equal the canonical energy associated with the region $R_B$ in in the spacetime $M(\lambda)$ . This statement does not make any additional assumptions beyond the HRT formula; in particular, it does not assume a full AdS/CFT correspondence.

In the context of a consistent theory of quantum gravity for asymptotically AdS spacetimes, the positivity of quantum Fisher information in the dual CFT implies that canonical energy for each region $R_B$ must be positive for physically consistent spacetimes. In the case of pure gravity, or specific examples of gravity coupled to matter, it should be possible to check this positivity explicitly for all allowed perturbations to AdS; partial results along these lines were given in \cite{blanco2013relative,Banerjee:2014oaa,Banerjee:2014ozp}. More generally, we can view these constraints as conditions on the stress-energy tensor that must be satisfied for any spacetime in any consistent theory. Specifically, equation (\ref{metricConstr1}) generalizes partial results for the energy condition arising from positivity of relative entropy at second order given in \cite{Lin:2014hva,Lashkari:2014kda}.  This condition is implied by the weak energy condition but is a weaker integrated version. The condition may be interpreted as requiring the positivity of Rindler energy for all possible wedges $R_B$.

The present work focuses on constraints on asymptotically AdS spacetimes at second order in perturbations around pure AdS arising from positivity of relative entropy. These can be viewed as a special case of a general set of constraints on arbitrary asymptotically AdS spacetimes from the monotonicity of relative entropy.\footnote{In this context, all positivity constraints follow from the monotonicity constraints.} In a forthcoming paper, we will describe how the technology of Hollands and Wald can be used to describe these most general relative entropy constraints as inequalities on bulk integrals involving the matter stress-energy tensor.

While the explicit examples in this paper have focused on Einstein gravity coupled to matter, the Wald formalism applies to general covariant theories of gravity. For these more general theories, the entanglement entropy formula must be generalized \cite{Camps:2013zua,Dong:2013qoa}, but we expect that all the main results carry over as they did in the case of the first order analysis \cite{faulkner2014gravitation}. It would also be interesting to extend the analysis in this work to the semiclassical level (as for the first-order analysis in \cite{swingle2014universality}), where the holographic entanglement entropy formula includes a contribution from entanglement entropy of bulk quantum fields \cite{Faulkner:2013ana}.

It would be interesting to understand the gravity interpretation of quantum Fisher information more generally, e.g. for perturbations around other solutions to Einstein equations. On the other hand, there are many other contexts  where canonical energy is well-defined, e.g. for perturbations to black holes in AdS or in more general spacetimes. It would be interesting to understand whether in these cases also canonical energy may be identified with Fisher information in some underlying quantum system. Assuming this to be the case might provide hints on the Hilbert space structure of the underlying quantum theory for cases where we currently do not have a nonperturbative description.

The identification of canonical energy with the Fisher information provides another link between quantum information theory and gravitational physics in the context of the AdS/CFT correspondence.\footnote{For other recent interesting examples of specific connections between natural concepts and quantities in quantum information and natural quantities in gravitational theories, see for example \cite{Czech:2014tva,Almheiri:2014lwa,MIyaji:2015mia,Stanford:2014jda,Bhattacharya:2014vja}.} Such identifications allow us to promote geometrical quantities which are well-defined in the classical (or semiclassical) limit of the gravity theory to quantities which are completely well-defined in the full quantum theory provided by the CFT dual. Making use of these identifications should help us to ask physical questions about gravity in a fully quantum-mechanical regime, beyond the semiclassical approximation.

\section*{Acknowlegements}

We thank Stefan Hollands, Hirosi Ooguri and Bob Wald for helpful conversations. We acknowledge the support of the KITP during the programs ``Entanglement in Strongly-Correlated Quantum Matter'' and ``Quantum Gravity: from UV to IR'' where some of this work was
done. The research of MVR and NL is supported in part by the Natural Sciences and Engineering Research Council of Canada and by FQXi. The work of MVR was supported by grant 376206 from the Simons Foundation.

\bibliographystyle{JHEP}

\bibliography{relative_constraint}

\end{document}